\documentclass[iop,apj,tighten]{emulateapj}
\usepackage{apjfonts} %If missing fonts, comment out this or google apjfonts to download
\usepackage{amsmath,amstext}
\usepackage{graphicx}
\usepackage{rotating}
\usepackage{multirow}
\usepackage{threeparttable}
\usepackage[breaklinks,colorlinks,citecolor=blue,linkcolor=magenta]{hyperref} 
\usepackage[all]{hypcap} %Links go to figures. This breaks deluxetables; use \capstartfalse \capstarttrue around deluxetables to fix it.

\begin{document}

\title{Energy release in the solar atmosphere from a stream of infalling prominence debris}
\author{A. R. Inglis\altaffilmark{1,2}, H. R. Gilbert\altaffilmark{1} and L. Ofman\altaffilmark{1,2}}
\affil{1. Solar Physics Laboratory, Heliophysics Science Division, NASA Goddard Space Flight Center, Greenbelt, MD, 20771}
\affil{2. Physics Department, The Catholic University of America, Washington, DC, 20064}

\begin{abstract}
Recent high-resolution and high-cadence EUV imaging has revealed a new phenomenon, impacting prominence debris, where prominence material from failed or partial eruptions can impact the lower atmosphere, releasing energy. We report a clear example of energy release and EUV brightening due to infalling prominence debris that occurred on 2011 September 7-8. The initial eruption of material was associated with an X1.8-class flare from AR11283, occurring at 22:30 UT on 2011 September 7. Subsequently, a semi-continuous stream of this material returned to the solar surface with a velocity v > 150 km/s, impacting a region remote from the original active region between 00:20 - 00:40 UT on 2011 September 8. Using SDO/AIA, the differential emission measure of the plasma was estimated throughout this brightening event. We found that the radiated energy of the impacted plasma was $L_{rad}$ $\sim$ 10$^{27}$ ergs, while the thermal energy peaked at $\sim$ 10$^{28}$ ergs. From this we were able to determine the mass content of the debris to be in the range 2 $\times$ 10$^{14}$ $<$ $m$ $<$ 2 $\times$ 10$^{15}$ g. Given typical promimence masses, the likely debris mass is towards the lower end of this range. This clear example of a prominence debris event shows that significant energy release takes place during these events, and that such impacts may be used as a novel diagnostic tool for investigating prominence material properties.
\end{abstract}

\keywords{Sun: filaments, prominences --- Sun: UV radiation --- Sun: corona --- Sun: activity}
\maketitle

\section{Introduction}
\label{introduction}

It has been known for some time that solar prominences, or filaments, exhibit a wide range of eruptive behaviour, up to and including the full ejection of significant material from the solar corona into the heliosphere. More commonly, either a partial or failed eruption is observed \citep{2007SoPh..245..287G}, where some or all of the eruptive prominence debris fails to escape the solar atmosphere and falls back towards the surface. These types of prominence eruptions are closely associated with coronal mass ejections and are key towards improving our understanding of CME initiation.

Despite decades of research, several properties of prominences remain poorly constrained \citep[see][for recent reviews]{2010SSRv..151..243L, 2010SSRv..151..333M}. For example, a novel method of determining prominence mass was presented by \citet{2005ApJ...618..524G}, yet the uncertainties in column density remained substantial. Additionally, the filling factor of prominences, as with many other solar phenomena, remains poorly known \citep{1998SoPh..183..107K}. In recent years however, increased availability of state-of-the-art solar instrumentation, including those on-board the Solar Dynamics Observatory (SDO), STEREO and IRIS, has dramatically increased the potential for detailed studies of this phenomenon.

In particular, recent observations have shown that descending prominence debris from failed eruptions can cause substantial energy release and plasma heating upon impact with the solar atmosphere \citep[e.g.][]{2013ApJ...776L..12G, 2013Sci...341..251R, 2014ApJ...797L...5R}. This energy release is directly observable by SDO at EUV wavelengths via the Atmospheric Imaging Assembly (AIA), providing a new diagnostic opportunity for understanding the properties of CME-associated material and probing the response of the solar atmosphere. The best example of this phenomenon observed to date is the flare- and CME-associated 2011 June 7 eruption \citep{2013Sci...341..251R, 2013ApJ...776L..12G, 2012A&A...540L..10I, 2016A&A...592A..17I, 2013ApJ...777...30I, 2014ApJ...782...87C, 2014ApJ...788...85V, 2016ApJ...827..151Y, 2012ApJ...746...13L}, where localized EUV brightening was observed at multiple impact points due to descending prominence debris. Such brightening patches are spatially and temporally resolved by SDO/AIA at multiple wavelengths, indicating that the plasma is multithermal and heated to several MK \citep{2013ApJ...776L..12G, 2013Sci...341..251R}. \citet{2013Sci...341..251R} compared these observations with the process of stellar accretion observed at UV and X-ray wavelengths. Recently, such EUV brightening was observed in another flare by \citet{2017ApJ...838...15L}.

Other phenomena sometimes compared to prominence debris include sequential chromospheric brightenings \citep[SCBs;][]{2005ApJ...630.1160B, 2012ApJ...750..145K, 2017SoPh..292...72K}, and coronal rain in active regions \citep{2012SoPh..280..457A, 2015ApJ...806...81A, 2015A&A...577A.136V}. Both these phenomena are substantially different; SCBs generally occur only in the parent active region of the eruption, while the small descending blobs associated with coronal rain have much lower downward speeds than eruptive prominence debris, do not cause observable emission due to impacting in the solar atmosphere, and are not associated with prominence or CME material. 

In this work, we present a new case study of prominence debris impacting the lower atmosphere, from 2011 September 7-8. To the best of our knowledge, this is only the second example of this phenomenon subject to detailed study, and the most energetic observed to date. This event took the form of a single, near-continuous material stream leading to continuous brightening of an atmospheric region at multiple wavelengths. By examining its energetic and kinematic properties we constrain the properties of the incoming stream, including the total mass of the deposited material. In Section \ref{obs} we describe the initial observations of this eruption, while in Section \ref{analysis} we present the methodology for estimating the radiated and thermal energy of the plasma, and infer the prominence mass. The implications of these estimates are discussed in Section \ref{discussion}.

\section{Observations}
\label{obs}

\begin{figure*}
\begin{center}
\includegraphics[width=10cm]{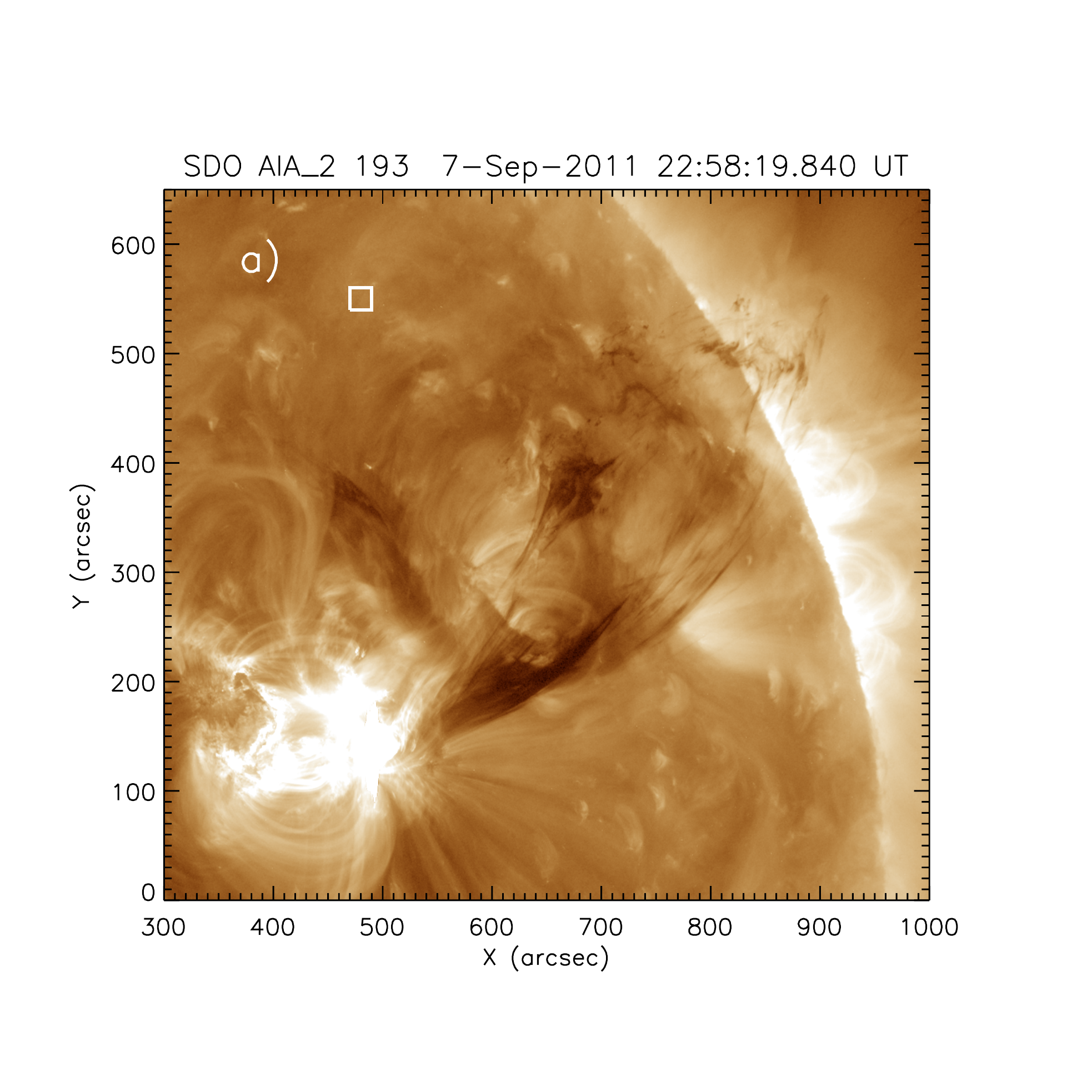}
\hspace{-1cm}
\includegraphics[width=7cm]{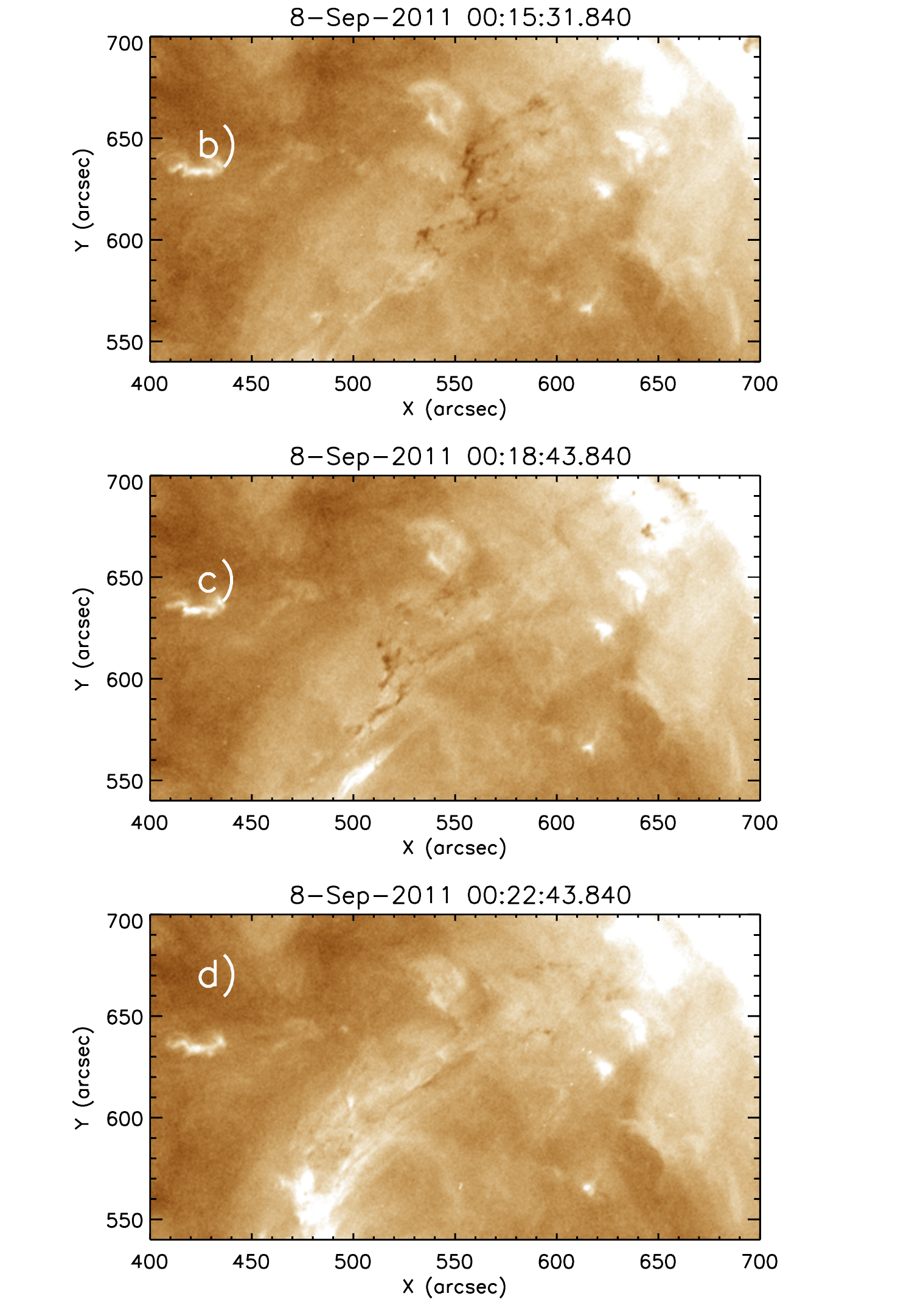}
\caption{The initial eruption beginning on 2011-09-07 is shown in panel a), where the white box indicates the impact point of the prominence debris. Panels b), c) and d) show the descending material stream over an hour later, at ~00:15 on 2011-09-08. The color contrast in panels b), c) and d) has been stretched to enhance the cool material.}
\label{summary_fig}
\end{center}
\end{figure*}

The initial prominence eruption occurred at $\sim$ 22:32 UT on 2011 September 7, as part of an X1.8 class solar flare originating from AR 11283 \citep[e.g.][]{2013JPhCS.440a2046Z}. This event was also associated with a non-geoeffective CME. Figure \ref{summary_fig}a shows the partially eruptive prominence material shortly after flare onset, at 22:58 UT. From this it is clear that a large amount of cool material was energized during the flare impulsive phase. However, despite the large energy release during this event a substantial amount of material fails to escape the Sun, and returns to the solar surface two hours after flare onset, at $\sim$ 00:20 UT on 2011 September 8. The impact point of the stream is substantially removed from the original active region and is indicated by the white box in Figure \ref{summary_fig}a. The main stream of material is shown in Figures \ref{summary_fig}b, c and d. These show successive snapshots of the stream just prior to impact with the lower corona. Figure \ref{summary_fig}d shows the beginning of the atmospheric brightening due to this material, indicating that the plasma is being substantially heated.

\section{Analysis and Results}
\label{analysis}

\subsection{Impact evolution}
\label{subsec_evolution}
\begin{figure*}
\begin{center}
\includegraphics[width=16cm]{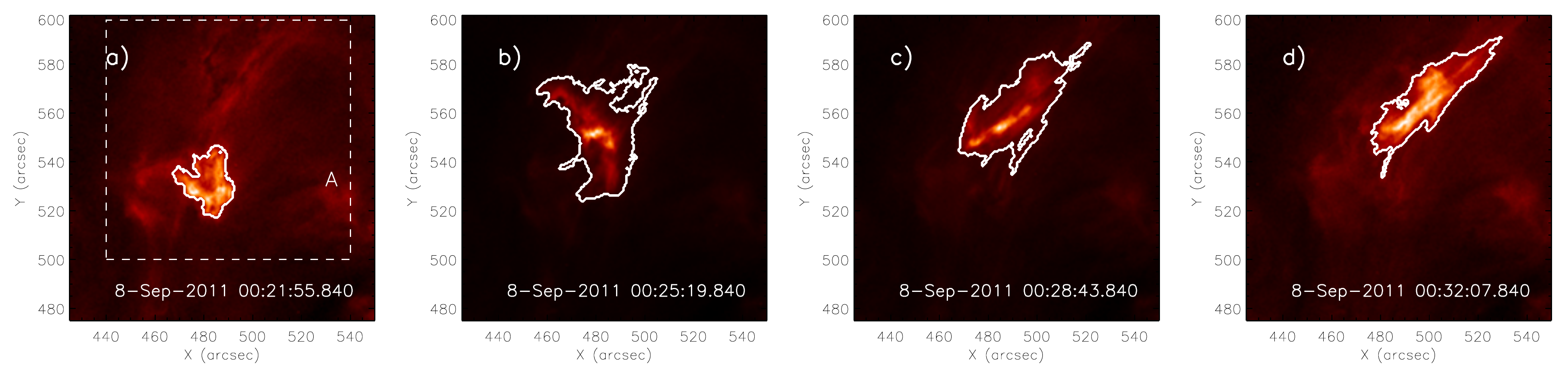}
\includegraphics[width=12cm]{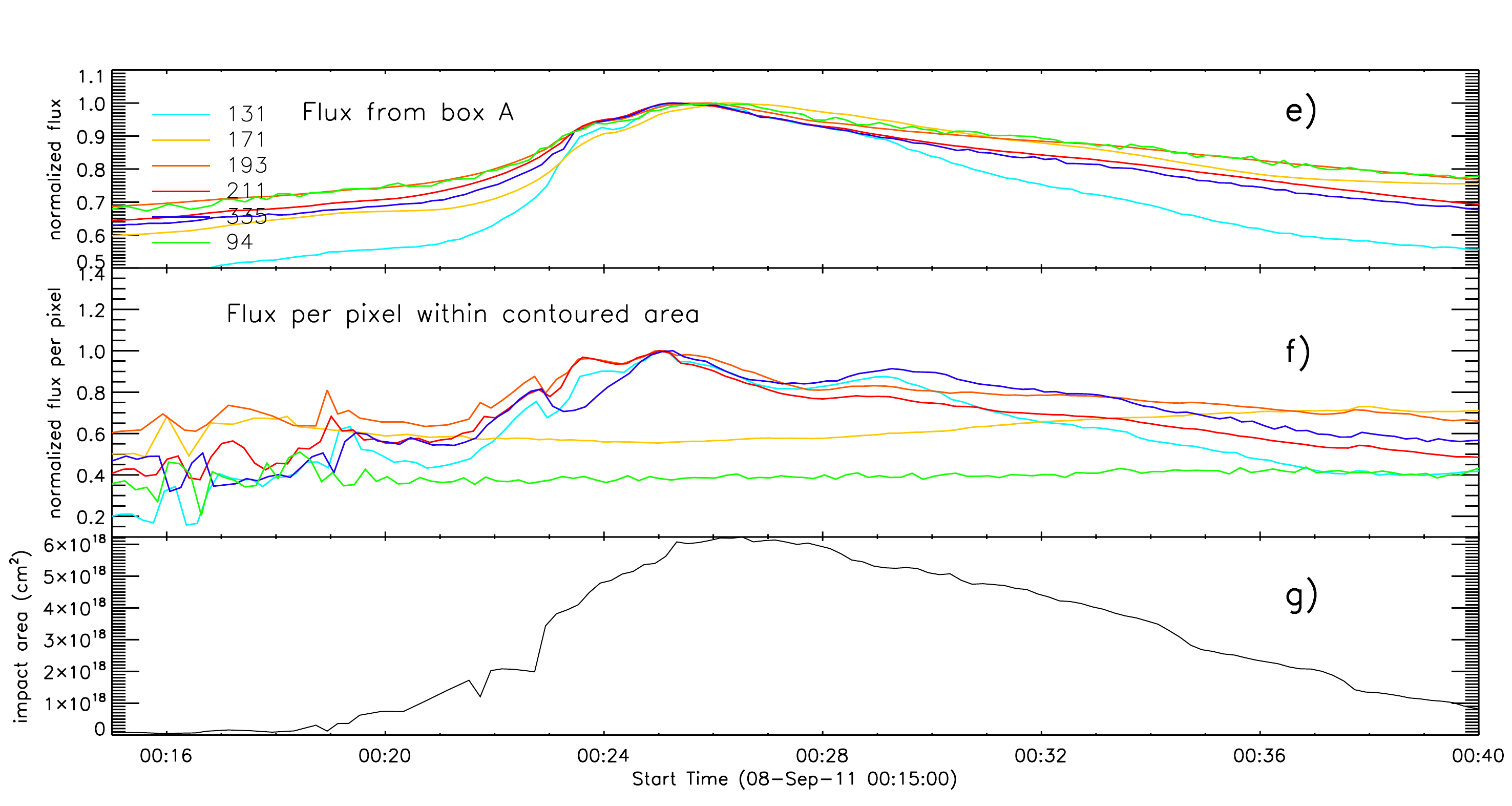}
\caption{Evolution of the bright impact area due to the prominence material stream. Panels a)-d) show the hot plasma observed by SDO/AIA at 193A at different times during impact. The white contour indicates the estimated area of the brightening at each time. Box A denotes a constant-area region used to illustrate the overall change in flux. Panel e) shows the integrated, normalized flux from box A in the 6 optically thin EUV channels. Panel f) shows the normalized, pixel-averaged lightcurves for the area contained within the contour. Panel g) shows the estimate of the brightening area during the event, as described in Section \ref{subsec_evolution}.}
\label{impact_evolution}
\end{center}
\end{figure*}

Figure \ref{impact_evolution} shows the evolution of the bright region caused by the impacting prominence stream as observed by SDO/AIA. Panels a)-d) show the appearance of the bright region at 4 different times. Clearly, the bright source undergoes substantial evolution over time during the stream impact. Hence, it is necessary to construct a scheme to estimate the area of the brightening over time. To achieve this we first find the point in space and time corresponding to the maximum brightening value in the SDO/AIA 193\AA\ channel. From this we establish a threshold of 5\% of this maximum for a pixel to be included in the bright region. Using this threshold, for each image frame we find the point of maximum intensity in the brightening region, and expand in all directions until the threshold is reached. Thus, for each frame we estimate the area of the instantaneous bright, heated plasma. The white contour in panels a)-d) shows the area defined by this scheme during the selected times. Figures \ref{impact_evolution}e-g show the full evolution of this brightening region over time. Figure \ref{impact_evolution}e shows the integrated flux from a constant area that encompasses the entire brightening (shown as Box A in panel a). This illustrates that all six EUV channels experience an increase in flux during impact. However, the 171\AA\ flux peaks slightly later than most other channels, reaching a maximum at $\sim$ 00:26 UT, compared with $\sim$ 00:25 UT for the 211\AA\ and 335\AA\ emission, consistent with cooling of the bright plasma following impact. Figure \ref{impact_evolution}f shows instead the pixel-averaged, normalized flux in each channel within the contoured region only. Although the majority of channels show an increase in flux on a per-pixel basis, the 94\AA\ and 171\AA\ channels are exceptions, indicating that the increased brightening in these channels is due almost entirely to the increasing size of the bright area, rather than increasing flux in each pixel. This is highlighted in Figure \ref{impact_evolution}g, which shows the estimated area in cm$^2$ of the bright region as it evolves. Together, these panels show that brightening begins at $\sim$ 00:20 UT, peaking at ~00:26 UT, with emission continuing until $\sim$ 00:40 UT. The impact area of the 2011 September 7-8 event is much larger than the brightenings observed on 2011 June 7; from Figure \ref{impact_evolution}f we see that the area peaks at $\sim$ 6 $\times$ 10$^{18}$cm$^2$. For comparison, the largest of the impacts observed on 2011 June 7 was $\sim$ 1.3 $\times$ 10$^{18}$ cm$^2$.

\begin{figure}
\begin{center}
\includegraphics[width=8.5cm]{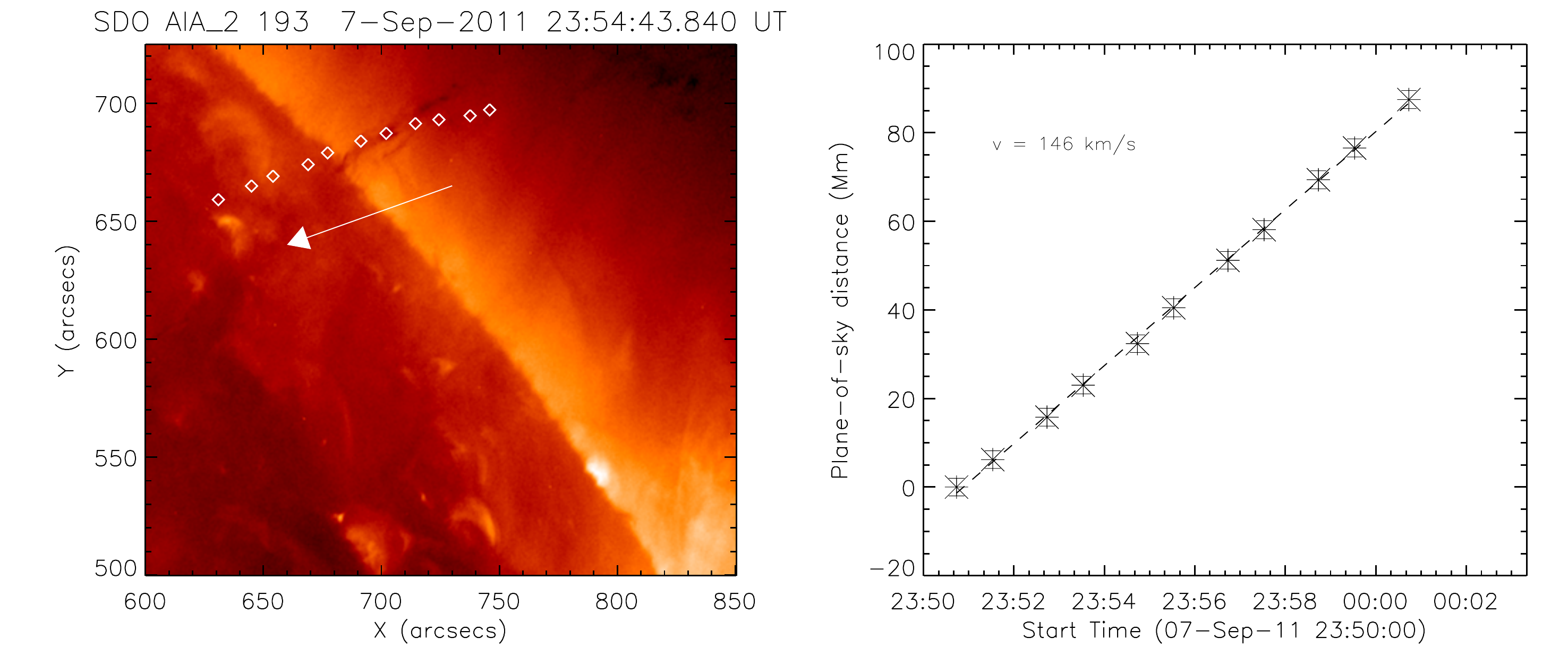}
\caption{Velocity estimate of the descending prominence debris stream. a) The location of a distinct piece of the material stream in successive AIA 193\AA\ image frames. b) Linear fit to the position estimates of the material, yielding a plane-of-sky velocity v = 146 km/s.}
\label{velocity_fig}
\end{center}
\end{figure}

The descending prominence debris is visible in several consecutive AIA image frames prior to impact with the lower corona, as shown in Figure \ref{summary_fig}. Thus, we can estimate the plane-of-sky velocity of the infalling material. Although the true observing angle of the material stream cannot be determined due to a lack of triangulation, plane-of-sky measurements can place lower limits on the stream velocity. In Figure \ref{velocity_fig}a, we estimate the position of a distinct piece of the descending stream in successive AIA 193\AA\ images. These locations are shown by the white diamonds, propagating from solar west to east. In Figure \ref{velocity_fig}b, we perform a linear fit to these positions, finding the best-fit plane-of-sky velocity v $\sim$ 150 km/s. However, there is large uncertainty due to projection effects; for example, if the material was actually propagating at 45$^{\circ}$ in the z-direction, the true velocity would be $\sim$ 220 km/s. Nevertheless, these values are similar to estimates of material velocity found by \citet{2013ApJ...776L..12G} for the 2011 June 7 event, who used triangulated measurements from AIA and STEREO-A, finding $v$ $\sim$ 150 - 300 km/s. For the same event, \citet{2013Sci...341..251R} estimated $v$ $\sim$ 300 - 450 km/s.  For the 2011 September 7-8 event, the material also does not appear to experience significant acceleration during this time period, suggesting it may have already reached critical velocity.

\subsection{Differential emission measure, energetics, and mass estimation}
Given the enhancements in emission from multiple SDO/AIA channels, we can investigate the energy release during the impact process by estimating the differential emission measure (DEM) of the bright plasma. To estimate the DEM, we use the forward fitting technique developed by \citet{2013SoPh..283....5A}, which was used by \citet{2013ApJ...776L..12G} to estimate the energy of prominence debris impacts in the 2011 June 7 event. We choose a DEM distribution of the form,

\begin{equation}
DEM(T) = EM_0 \exp \left(\frac{\log T - \log T_c}{2\sigma^2} \right)
\end{equation}

i.e. a Gaussian emission measure distribution with peak temperature $T_c$ and width $\sigma$, as utilized by \citet{2013SoPh..283....5A, 2015ApJ...802...53A}.

The temperature response functions of the AIA channels are the source of significant uncertainty \citep[e.g.][]{2013SoPh..283....5A}, particularly the 94\AA\ and 131\AA\ channels at low temperatures. To account for this, we include a 25\% uncertainty in the measured AIA flux due to instrument response, as suggested by \citet{2012SoPh..275...41B, 2012ApJS..203...25G, 2012ApJS..203...26G}. This is combined in quadrature with the statistical uncertainty associated with the AIA flux measurements.

\begin{figure}
\begin{center}
\includegraphics[width=8cm]{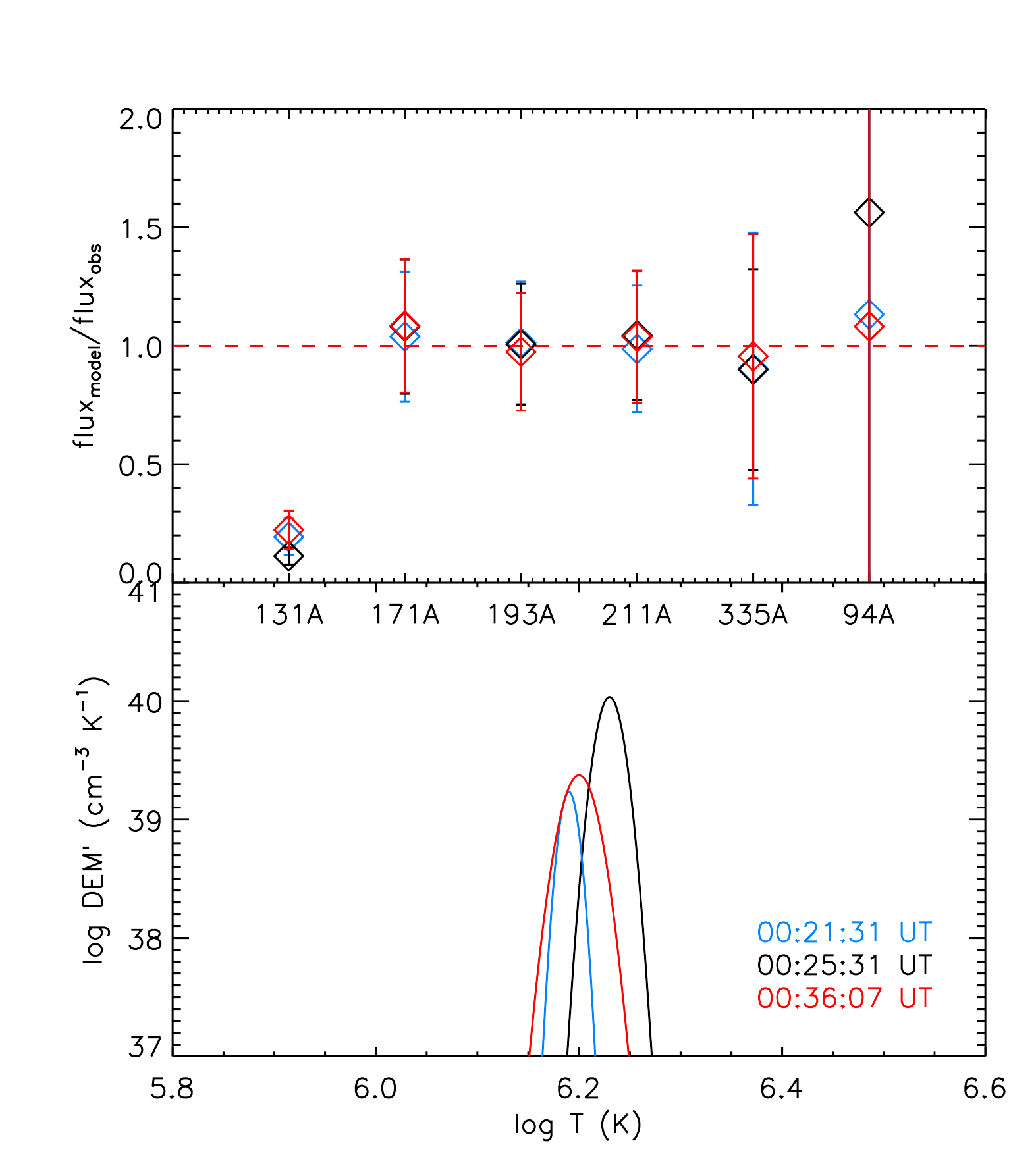}
\caption{Examples of the DEM fits to the AIA flux data, at three different times; 00:21:31 UT (blue), 00:25:31 UT (black), and 00:36:07 UT (red). The top panel shows the ratio of the modelled AIA flux to the observed flux in each channel. The bottom panel shows the best-fit Gaussian DEM functions for the three example times.}
\label{example_dem}
\end{center}
\end{figure}

The best fit to the observed flux is achieved at each time interval via a search over the parameter space given by the variables $EM_0$, $T_c$ and $\sigma$ using the $\chi^2$ test. Figure \ref{example_dem} shows examples of the best-fit DEM results at three different times, near the start, peak and end of the brightening. This shows that the majority of the brightening comes from increased $EM$ at moderate temperatures, with $\log T_{c} <$ 6.4. 

Given a DEM, we can estimate the radiative losses from the plasma via \citep[e.g.][]{2005psci.book.....A}, 

\begin{equation}
\frac{dL_{rad}}{dt} = \int^{T_2}_{T_1} DEM(T)^{\prime} \times \Lambda(T) \ dT \ \text{erg s}^{-1}
\label{lrad_eqn}
\end{equation}

where $\Lambda(T)$ is the radiative loss function and $DEM(T)^{\prime} = DEM(T) \times A$ is the differential emission measure multiplied by the emitting area $A$, and hence is in units of cm$^{-3}$ K$^{-1}$. Here $\log T_1$ = 5.5 and $\log T_2$ = 7.0. To find the total energy radiated, we estimate $\Lambda(T)$ using the CHIANTI database \citep{2012ApJ...744...99L, 2015A&A...582A..56D} assuming coronal abundances, and integrate Equation \ref{lrad_eqn} over the duration of the impacting stream, hence,

\begin{equation}
L_{rad} = \int^{t_1}_{t_0} \frac{dL_{rad}(t)}{dt} dt,
\end{equation}

where $t_0$ = 00:20 UT and $t_1$ = 00:40 UT.

Using the DEM, it is also possible to calculate the peak thermal energy $U_{th}$. For an isothermal plasma, this is given by \citep[e.g.][]{2005ApJ...621..482V, 2008ApJ...677..704H, 2012ApJ...759...71E, 2014ApJ...789..116I, 2016A&A...588A.116W},

\begin{equation}
U_{th} = 3 k_B T \sqrt{EM_{tot} f V}
\label{isothermal}
\end{equation}

where $EM_{tot}$ is the total emission measure in cm$^{-3}$ of the plasma with single temperature $T$, $V$ is the plasma volume and $f$ is the plasma filling factor. For a multi-thermal plasma, we must account for the energy at all $T$. Hence, Equation \ref{isothermal} becomes \citep{2014ApJ...789..116I, 2015ApJ...802...53A},

\begin{equation}
U_{th} = 3 k_B V^{1/2} \frac{\int^T DEM(T)^{\prime} \times T dT}{EM_{tot}^{1/2}}
\label{multithermal}
\end{equation}

where $DEM(T)^{\prime}$ is differential emission measure expressed in cm$^{-3}$ K$^{-1}$ as before, and a filling factor $f$ = 1 is assumed. The temperature bounds are the same as in Equation \ref{lrad_eqn}. For both Eqn \ref{isothermal} and \ref{multithermal} the volume $V$ must be estimated, which is complicated by a lack of observational information of the source in the z-direction. In this work, we use the simple estimate $V$ $\sim$ $A^{3/2}$, hence the estimated volume varies over time with the area (see Figure \ref{impact_evolution}).

\begin{figure}
\begin{center}
\includegraphics[width=8.5cm]{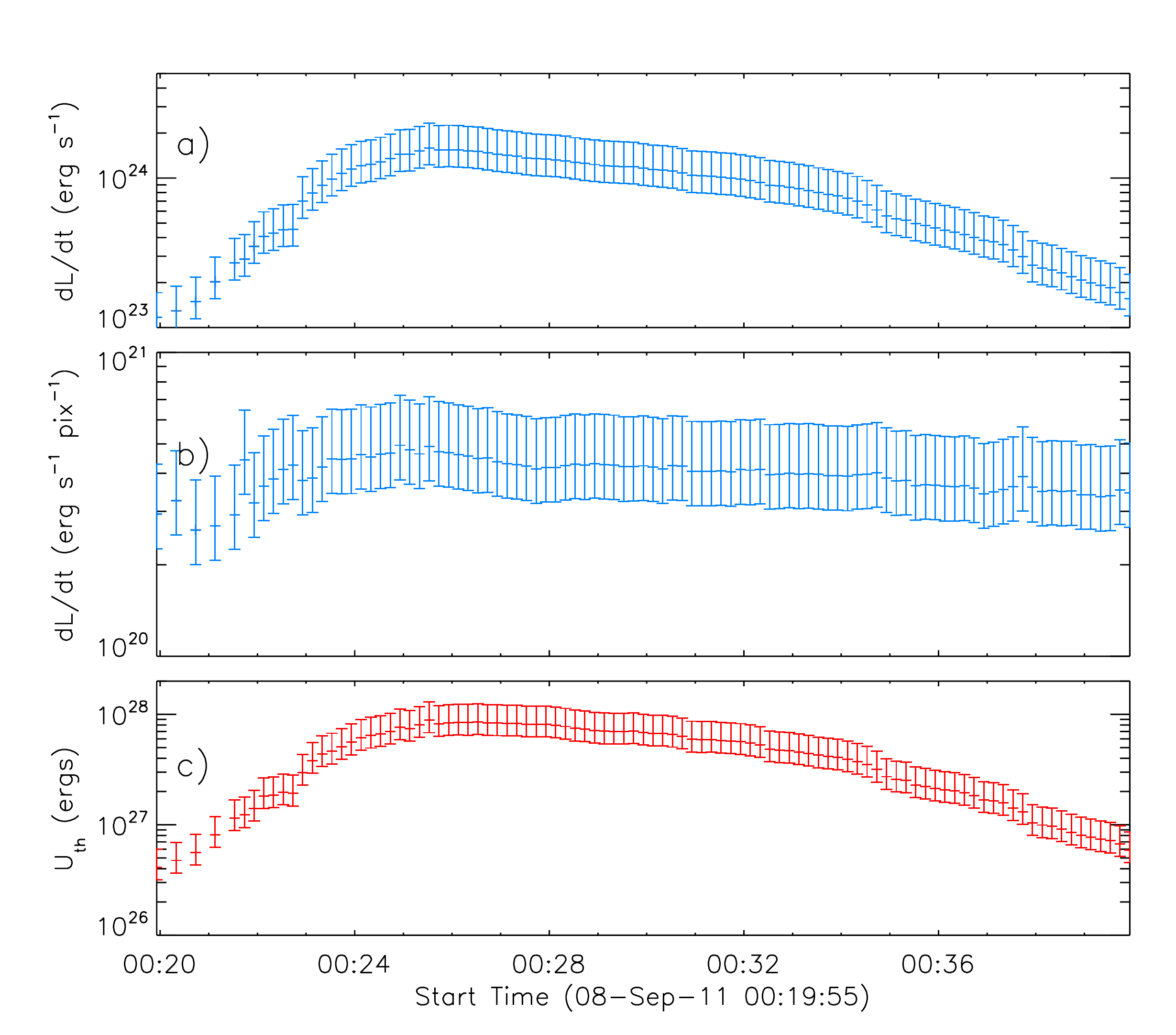}
\caption{Energetic properties of the impact region shown in Figure \ref{impact_evolution}. a) The radiated energy rate $dL/dt$ of the heated plasma as a function of time during the impacting stream. b) The radiated energy rate $dL/dt$ normalized to a per-pixel basis. c) The instantaneous thermal energy of the evolving impact plasma.}
\label{energetics_figure}
\end{center}
\end{figure}

Figure \ref{energetics_figure} shows the energetic properties of the debris impact region, the same region illustrated by the white contours in Figure \ref{impact_evolution}. Panel a) illustrates the estimated radiated energy rate $dL/dt$ as a function of time. The radiated losses show an order of magnitude increase, beginning at ~00:20 UT. This coincides with a substantial increase in the estimated area of the impact region (see Figure \ref{impact_evolution}); on a per-pixel basis, the increase in emission is smaller, approximately a factor $\sim$ 2. Hence, as Figure \ref{energetics_figure}b shows, the increase in radiated energy is a combination of increased flux and the enlargement of the emitting region itself. Figure \ref{energetics_figure}c shows the estimated instantaneous thermal energy during this event, peaking at $\sim$ 10$^{28}$ ergs. 

We can compare these energetic properties with the previous observations of impacting prominence debris from the 2011 June 7 event \citep{2013ApJ...776L..12G, 2013Sci...341..251R}. In \citet{2013ApJ...776L..12G}, the radiated energy was estimated for 5 observable brightening regions affected by impacting debris. Combining all of these regions, the estimated total radiated energy $L_{rad}$ was $\sim$ 4 $\times$ 10$^{26}$ ergs. For the 2011 September 7-8 observation, we find from integrating Figure \ref{energetics_figure}a that the total energy radiated is $\sim$ 10$^{27}$ ergs, at least a factor of 2 higher. This is consistent with the observations for two reasons; firstly, the 2011 September 7-8 observation consists of a relatively continuous material stream that impacts a larger area than the 2011 June 7 impacts, and secondly the brightening duration is substantially longer, with significant emission lasting for $\sim$ 20 minutes. 

However, the total estimated radiated energy loss is an order of magnitude lower than estimated peak thermal energy of the plasma $U_{th}$. This could be the result of two major factors. Firstly, when calculating the thermal energy, a value of $f$ $\sim$ 1 was assumed for the plasma filling factor, however, the true value of $f$ is unknown, and may be substantially less than unity \citep[e.g.][]{1997ApJ...478..799C, 2012ApJ...755...32G}. This, combined with uncertainties in the plasma volume $V$, leads to a large uncertainty in the estimate of $U_{th}$, possibly an overestimate. The second factor is that conductive losses could play an important role in the energy evolution of the bright plasma. 

Despite these uncertainties, we can use these energy estimates to constrain the minimum and maximum kinetic energy, and thus the mass, of the impacting prominence material. We can assume that the estimate of $L_{rad}$ gives us a lower limit on the kinetic energy requirement from the infalling prominence material. Hence $KE$ $\geq$ 10$^{27}$ ergs. Given our plane-of-sky velocity estimate of the falling material of v $\sim$ 150 km/s, we can estimate the minimum value of mass $m$ required to produce this kinetic energy. From this we estimate $m_{low}$ $\sim$ 2 $\times$ 10$^{14}$ g. Alternatively, we can assume that conductive losses play a significant role in the energy budget of this event, and that the estimate of $U_{th}$ provides a good approximation of the total energy deposited by the prominence material. In this case, we find that $KE$ $\sim$ 10$^{28}$ ergs, which requires $m_{high}$ $\sim$ 2 $\times$ 10$^{15}$ g. 

\section{Conclusions}
\label{discussion}

We have analysed a new example of energy release and EUV brightening due to prominence debris, occurring on 2011 September 8 at $\sim$ 00:20 UT. This event followed the large X-class flare from AR 11283 beginning at $\sim$ 22:30 UT on 2011 September 7. During the eruption, a large amount of cool material was ejected from the active region. Some of that material failed to escape the solar atmosphere, resulting in a descending stream of prominence debris that impacted the atmosphere at a different location from the original AR. This caused an extended EUV brightening lasting for $\sim$ 20 minutes between 00:20 UT and 00:40 UT. 

Due to the brightening of emission in multiple optically thin AIA channels during impact, we estimated the DEM of the plasma as a function of time throughout the event, and consequently the peak thermal energy and radiated energy of the plasma (see Figure \ref{energetics_figure}). We found the peak value of thermal energy to be $U_{th}$ $\sim$ 10$^{28}$ ergs. The estimated total radiated energy was an order of magnitude smaller, at $L_{rad}$ $\sim$ 10$^{27}$ ergs. The disparity between these values may be due either to the importance of conductive losses in the plasma, or the uncertainty in the plasma volume $V$ and the filling factor $f$, both of which effect the estimate of $U_{th}$. Comparing these values to estimates from the well-known 2011 June 7 event \citep[e.g.][]{2013ApJ...776L..12G, 2013Sci...341..251R}, we see that the 2011 September 7-8 event was more energetic overall, lasting for longer and radiating more energy. 

Using SDO/AIA images we also estimated the plane-of-sky velocity of the descending stream at $v$ $\sim$ 150 km/s. Using this as a lower limit on the true velocity, we constrain the kinetic energy and mass requirements of the prominence material in order to explain the observed brightening. We estimate that  10$^{27}$ $<$ $KE$ $<$ 10$^{28}$ ergs, and 2 $\times$ 10$^{14}$ $<$ $m$ $<$ 2 $\times$ 10$^{15}$ g. This is substantially greater than the estimate of mass obtained for the prominence debris from 2011 June 7 \citep{2013ApJ...776L..12G}, consistent with the larger energy release from this event.

We can compare this mass range estimate with the typical mass values expected from solar prominences, which range from 5 $\times$ 10$^{12}$ - 10$^{15}$ g \citep{2010SSRv..151..243L}, based on a wide range of possible prominence volumes. Similar estimates of $m$ $\sim$ 10$^{14}$ - 2 $\times$ 10$^{15}$ g were found by \citet{2006ApJ...641..606G}. Given these estimates, a debris mass of 10$^{15}$ g seems unphysical, as this approaches the mass of an entire prominence. Hence, the value of $m$ derived from $U_{th}$ for this event may be an overestimate due to uncertainty on the plasma filling factor $f$ and volume $V$, which introduces large uncertainty into the estimate of $U_{th}$ (see Equations \ref{isothermal}, \ref{multithermal}). The radiated energy estimate however does not depend on the plasma volume, and may provide a more realistic estimate of the impacting material. Another uncertain factor is the velocity $v$; the true value could be higher than the estimated plane-of-sky velocity, lowering the mass required to explain the observations. Hence, we conclude that the mass of impacting prominence debris is of order $m$ $\sim$ 2 $\times$ 10$^{14}$g.

This observation shows that prominence debris from partial or failed eruptions can cause significant energy release in the lower corona, at sites far removed from the initial active region. These observations can constrain the properties of the local coronal plasma and the prominence debris itself. Further study of this type of phenomenon would help to better constrain the energetic and kinematic properties of these events.

\bibliographystyle{apj}

\end{document}